\documentclass[a4paper,10pt,aps,prd,floatfix,notitlepage,showpacs]{revtex4-1}
\usepackage{amsmath,amssymb,amsfonts}

\usepackage{tikz}
\usetikzlibrary{matrix}

\DeclareMathOperator{\arcsinh}{arcsinh}

\begin{document}

\title{Modified teleparallel theories of gravity -- Gauss-Bonnet and trace extensions}

\author{Sebastian Bahamonde}
\email{sebastian.beltran.14@ucl.ac.uk}

\author{Christian G. B\"ohmer}
\email{c.boehmer@ucl.ac.uk}

\affiliation{Department of Mathematics, University College London, Gower Street, London, WC1E 6BT, United Kingdom}

\date{\today}

\begin{abstract}
  We investigate modified theories of gravity in the context of teleparallel geometries with possible Gauss-Bonnet contributions. The possible coupling of gravity with the trace of the energy-momentum tensor is also taken into account. This is motivated by the various different theories formulated in the teleparallel approach and the metric approach without discussing the exact relationship between them. Our formulation clarifies the connections between different well known theories. For instance, we are able to formulate the correct teleparallel equivalent of Gauss-Bonnet modified general relativity, amongst other results. Finally, we are able to identify modified gravity models which have not been studied in the past. These appear naturally within our setup and would make a interesting starting point for further studies.
\end{abstract}

\maketitle

\section{Introduction}

One possible approach to motivating a geometrical theory of gravity is to compare the geodesic equation of differential geometry with Newton's force law. This suggests the identification of the gravitational forces with the components of the Christoffel symbols which in turn yields the identification of the gravitational potentials with the metric. Assuming a geometrical framework with a metric compatible covariant derivative without torsion gives the building block of Einstein's theory of general relativity, one can speak of the metric approach.

Soon after the original formulation of this geometrical theory of gravity, it was noted that there exits an alternative geometrical formulation which is based on a globally flat geometry with torsion. The key mathematical result to this approach goes back to Weitzenb\"ock who noted that it is indeed possible to choose a connection such that the curvature vanishes everywhere. This formulation gives equivalent field equations to those of general relativity and we refer to this as the teleparallel formulation. This naming convention stems from the fact that the notion of parallelism is global instead of local on flat manifolds, see for instance \cite{Aldrovandi:2013wha,Maluf:2013gaa} and reference therein. 

One of the basic equations of general relativity and its teleparallel equivalent is $R = -T + B$ where $R$ is the Ricci scalar, $T$ is the torsion scalar and $B$ is a total derivative term which only depends on torsion. Clearly, the Einstein-Hilbert action can now be represented in two distinct ways, either using the Ricci scalar or the torsion scalar, and consequently giving identical equations of motion. A popular modification of general relativity is based on the Lagrangian $f(R)$ and can be viewed as a natural non-linear extension that results in fourth order field equations \cite{Sotiriou:2008rp,DeFelice:2010aj}. On the other and, one could consider the Lagrangian $f(T)$ which gives second order field equations \cite{Ferraro:2006jd}. However, this theory is no longer invariant under local Lorentz transformations because the torsion scalar $T$ itself is not invariant \cite{Li:2010cg,Sotiriou:2010mv}. Neither is the total derivative term $B$ but the particular combination $-T + B$ is the unique locally Lorentz invariant choice, see also~\cite{Bahamonde:2015zma}. Hence, $f(R)$ gravity and $f(T)$ gravity and not equivalent and correspond to physically different theories. Now, considering the more general family of theories based on $f(T,B)$ one can establish the precise relationship between these theories and it turns out that $f(R)$ gravity is the unique locally Lorentz invariant theory while $f(T)$ gravity is the unique second order theory.

The principal aim of this paper is to extend these results to take into account the Gauss-Bonnet term and its teleparallel equivalent. The Gauss-Bonnet scalar is one of the so-called Lovelock scalars \cite{Lovelock:1971yv} which only yields second order field equations in the metric, hence in more than four dimensions the study of the Gauss-Bonnet term is quite natural. In four dimensions, on the other hand, the Gauss-Bonnet term can be written as total derivative and its integral over the manifold is related to the topological Euler number. However, it should be emphasised that topological issues in teleparallel theories are not well understood yet. The only known torsional topological invariant is the Nieh-Yan term, see \cite{Nieh:2007zz, Chandia:1997hu}, and in particular \cite{Mielke:2009zz} in the context of teleparallel theories.

The teleparallel equivalent of the Gauss-Bonnet term was first considered in \cite{Kofinas:2014owa,Kofinas:2014aka} who studied a theory based on the function $f(T,T_G)$. As is somewhat expected, the Gauss-Bonnet term differs from its teleparallel equivalent by a divergence term. Hence, as in modified general relativity, it is possible to formulate modified theories based on the Gauss-Bonnet terms or its teleparallel equivalent in such a way that both theories are physically distinct. The link between these theories comes from the divergence term which needs to be taken into account when establishing the relationship between the different possible theories. We also allow our action functional to depend on the trace of the energy-momentum tensor since this is a popular modification that has been investigated in recent years \cite{Harko:2011kv}.

There exits a large body of literature dealing with the various modified theories of gravity that have been investigated, see for instance the following reviews \cite{Sotiriou:2008rp,DeFelice:2010aj,Nojiri:2010wj,Capozziello:2011et,Cai:2015emx}

{\it Our conventions:} Greek indices denote spacetime coordinates, Latin indices are frame or tangent space indices. $e_{\mu}^a$ stands for the tetrad (1-form), while $E^{\mu}_a$ denotes the inverse tetrad (vector field). The Minkowski metric is $\eta_{ab}$ with signature $(-,+,+,+)$. Where possible we follow the conventions of~\cite{Maluf:2013gaa}.

\section{Teleparallel gravity and the Gauss-Bonnet term}
\label{sec:tegr}

\subsection{Teleparallel geometries}

The teleparallel formulation of general relativity is well known and provides some interesting insight into this theory. The fundamental objects of this formulation are the tetrad fields $e^{a}_\mu$ and  inverse tetrad fields $E^{\mu}_{a}$ which satisfy the orthogonality relations
\begin{align}
  E_{m}^{\mu} e_{\mu}^{n} &= \delta^{n}_{m} \,,
  \label{deltanm} \\
  E_{m}^{\nu} e_{\mu}^{m} &= \delta^{\nu}_{\mu} \,.
  \label{deltamunu}
\end{align}

The standard (metric) formulation of general relativity is based on the metric tensor which is uniquely defined by given tetrad fields. The metric $g_{\mu\nu}$, the inverse metric $g^{\mu\nu}$, and the tetrads and inverse tetrads are related by
\begin{align}
  g_{\mu\nu} &= e^{a}_{\mu} e^{b}_{\nu} \eta_{ab} \,, \label{metric}\\
  g^{\mu\nu} &= E^{\mu}_{a} E^{\nu}_{b} \eta^{ab} \,.
\end{align} 
The determinant of the tetrad $e^a_\mu$ is denoted by $e = \det (e^a_\mu)$ and corresponds to the volume element of the metric which means we have $e=\sqrt{-g}$ where $g = \det (g_{\mu\nu})$ is the determinant of the metric tensor.

The starting point of the teleparallel formulation of general relativity is the object
\begin{align}
  W_{\mu}{}^{a}{}_{\nu} = \partial_{\mu}e^{a}{}_{\nu} \,,
\end{align}
from which we can define the torsion tensor
\begin{align}
  T^{a}{}_{\mu\nu} &= W_{\mu}{}^{a}{}_{\nu} - W_{\nu}{}^{a}{}_{\mu} =
  \partial_{\mu} e_{\nu}^{a} - \partial_{\nu}e_{\mu}^{a} \,.
  \label{eq:tor}
\end{align}
We note that this is the skew-symmetric part of $W_{\mu}{}^{a}{}_{\nu}$. In terms of spacetime indices the torsion tensor is 
\begin{align}
  T^{\lambda}{}_{\mu\nu} = E_{a}^{\lambda} T^{a}{}_{\mu\nu} \,.
\end{align}
In geometries with torsion the connection can be decomposed into a Levi-Civita part and an additional part due to the presence of torsion. The complete connection which we call the Weitzenb\"ock connection decomposes as follows
\begin{align}
  W_{\lambda}{}^{\mu}{}_{\rho} = {}^0 \Gamma^{\mu}_{\lambda\rho} + K_{\lambda}{}^{\mu}{}_{\rho} \,,
  \label{eq:connection}
\end{align}
where ${}^0 \Gamma$ denotes the Levi-Civita connection and $K$ is the contortion tensor. The contortion tensor can  be expressed using the torsion tensor as follows
\begin{align}
  2K_{\mu}{}^{\lambda}{}_{\nu} = T^{\lambda}{}_{\mu\nu} - T_{\nu\mu}{}^{\lambda} +T_{\mu}{}^{\lambda}{}_{\nu}\,.\label{KT}
\end{align}
We note that the contortion tensor $K_{\lambda}{}^{\mu}{}_{\rho}$ is antisymmetric in its last two indices, this follows from the skew-symmetry of the torsion tensor in its last two indices, see (\ref{eq:tor}). Contracting the torsion tensor over the first and second index gives the so-called torsion vector
\begin{align}
  T_{\mu} = T^{\lambda}{}_{\lambda\mu} \,.
\end{align}

Considering a globally flat manifold means that the Riemann tensor vanishes identically $R_{abcs}(W) = 0$, which is always possible by choosing the Weitzenb\"ock connection. Next, we can decompose the Riemann tensor into a Levi-Civita part and another part due to torsion. The Levi-Civita part of this tensor can then be expressed in terms of the contorsion tensor as follows
\begin{align}
  ^{0}R^{\lambda}\,_{\mu\sigma\nu} = {}^0\nabla_{\nu}K_{\sigma}{}^{\lambda}{}_{\mu} - 
  {}^0\nabla_{\sigma}K_{\nu}{}^{\lambda}{}_{\mu} +
  K_{\sigma}{}^{\rho}{}_{\mu}K_{\nu}{}^{\lambda}{}_{\rho} -
  K_{\sigma}{}^{\lambda}{}_{\rho}K_{\nu}{}^{\rho}{}_{\mu} \,,
  \label{relation}
\end{align} 
where ${}^0\nabla$ stands for the covariant derivative with respect to the Leci-Civita connection. If we contract the first and third index of the Riemann tensor, we obtain the decomposition of the Ricci tensor
\begin{align}
  ^{0}R_{\mu\nu} = {}^0\nabla_{\nu}K_{\lambda}{}^{\lambda}{}_{\mu} - 
  {}^0\nabla_{\lambda}K_{\nu}{}^{\lambda}{}_{\mu} +
  K_{\lambda}{}^{\rho}{}_{\mu}K_{\nu}{}^{\lambda}{}_{\rho} -
  K_{\lambda}{}^{\lambda}{}_{\rho}K_{\nu}{}^{\rho}{}_{\mu} \,.
  \label{riccitensorsplit}
\end{align} 
Finally, by  using (\ref{KT}) and contracting once more, one arrives at the well-known formula 
\begin{align}
  R(e) + \left(\frac{1}{4}T^{abc}T_{abc}+\frac{1}{2}T^{abc}T_{bac}-T^aT_a\right) - \frac{2}{e} \partial_\mu (e T^\mu) = 0\,.
  \label{ricci0}
\end{align}
Here $R(e)$ stands for the metric or Levi-Civita Ricci scalar, see also~\cite{Maluf:2013gaa}. 

\subsection{Teleparallel gravity}

The main implication of the previous rewriting is that it is hence possible to express the metric Ricci scalar entirely in terms of torsion
\begin{align}
  R(e) = - \left(\frac{1}{4}T^{abc}T_{abc}+\frac{1}{2}T^{abc}T_{bac}-T^aT_a\right) + \frac{2}{e} \partial_\mu (e T^\mu) \,.
  \label{ricci1}
\end{align}
This result can also be written in the following way
\begin{align}
  R(e) = - S^{abc}T_{abc} + \frac{2}{e}\partial_\mu (e T^\mu) \,,
  \label{ricciS}
\end{align}
where the tensor $S^{abc}$ is defined as follows
\begin{align}
  S^{abc} = \frac{1}{4}(T^{abc}-T^{bac}-T^{cab})+\frac{1}{2}(\eta^{ac}T^b-\eta^{ab}T^c) \,.
\end{align}
Its form in spacetime coordinates can be written as
\begin{align}
  2S_{\sigma}{}^{\mu\nu} = K_{\sigma}{}^{\mu\nu} - \delta^{\mu}_{\sigma}T^{\nu} + 
  \delta^{\nu}_{\sigma}T^{\mu} \,.
  \label{S}
\end{align}
Frequently the specific combination $S^{abc}T_{abc}$ is referred to simply as the torsion scalar $T$ so that Eq.~(\ref{ricciS}) can be written very nicely as
\begin{align}
  R(e) = - T + \frac{2}{e}\partial_\mu (e T^\mu) \,.
  \label{ricciT}
\end{align}
Recalling that the Einstein-Hilbert action of General Relativity is based on the Ricci scalar, we are now able to formulate a theory equivalent to this based on torsion by simply using the right-hand side of (\ref{ricciT}) as our Lagrangian instead of the left-hand side. In either case one considers variations with respect to the tetrad fields. Since we will study the boundary term in some detail we introduce the notation
\begin{align}
  B = \frac{2}{e}\partial_\mu (e T^\mu) = 2 \nabla_\mu T^\mu \,,
  \label{B}
\end{align}
so that (\ref{ricciT}) can simply be written as $R = -T + B$. 

An important consideration is the behaviour of the above quantities under local Lorentz transformations. It is clear from (\ref{eq:tor}) that the local Lorentz transformation $e^a_\mu \mapsto \Lambda^a{}_b e^b_\mu$ will change the torsion tensor as the Lorentz transformations $\Lambda^a{}_b$ are local and hence functions of space and time so that derivatives of $\Lambda^a{}_b$ appear. Therefore the torsion tensor does not transform covariantly under local Lorentz transformations, see also \cite{Li:2010cg,Sotiriou:2010mv}. Note that this is a direct consequence of the teleparallel approach and the combination $-T+B$ is the only combination of $T$ and $B$ which is locally Lorentz invariant.

In contrast to the standard teleparallel approach, complete Lorentz invariance is preserved when considering metric-affine theories \cite{Hehl:1994ue} in which the metric and torsion and are treated independently, see also \cite{Obukhov:2006sk}. Consequently, the torsion scalar $T$ and the boundary term $B$ are both Lorentz scalars in this approach. The metric-affine framework has inspired the recent covariant formulation of $f(T)$ gravity \cite{Krssak:2015oua} which is based on the idea of allowing the spin connection to be a dynamical variable in addition to the tetrad fields. This is an interesting alternative treatment to teleparallel theories of gravity which could in also be applied to investigate Gauss-Bonnet extensions. 

\subsection{Gauss-Bonnet term}

Teleparallel geometries have been well-understood for many decades. Perhaps more surprising is the fact that the Gauss-Bonnet term was not studied in this context until quite recently~\cite{Kofinas:2014owa}. The Gauss-Bonnet term is a quadratic combination of the Riemann tensor and its contractions given by
\begin{align}
  G = R^2 - 4R_{\mu\nu}R^{\mu\nu} + R_{\mu\nu\kappa\lambda}R^{\mu\nu\kappa\lambda} \,,
  \label{G}
\end{align}
which plays an important role of connecting geometry to topology. It is well known that the addition of the Gauss-Bonnet to the Einstein-Hilbert action does not affect the field equations of general relativity, provided one works in a four dimensional setting. This fact implies that the topology of the solutions is unconstrained. In more than four dimensions the addition of the Gauss-Bonnet term affects the resulting gravitational field equations.  

Following the procedure outlined in the above, we can again compute the (complete) Gauss-Bonnet term using the connection (\ref{eq:connection}) and decompose this result into a Levi-Civita part and an additional part depending on torsion only. Understandably, this process is quite involved. It can be shown, see~\cite{Kofinas:2014owa,Gonzalez:2015sha}, that the Gauss-Bonnet term can be expressed in a fashion similar form to (\ref{ricciT}) which simply reads
\begin{align}
  G = -T_{G} + B_{G} \,.
  \label{GB}
\end{align}
The teleparallel Gauss-Bonnet term $T_{G}$ is given by
\begin{align}
  T_{G} = (
  K_{a}{}^{i}{}_{e}K_{b}{}^{ej}K_{c}{}^{k}{}_{f}K_{d}{}^{fl} -
  2K_{a}{}^{ij}K_{b}{}^{k}{}_{e}K_{c}{}^{e}{}_{f}K_{d}{}^{fl} +
  2K_{a}{}^{ij}K_{b}{}^{k}{}_{e}K_{f}{}^{el}K_{d}{}^{f}{}_{c} + 
  2K_{a}{}^{ij}K_{b}{}^{k}{}_{e}K_{c,d}{}^{el}
  )\delta^{abcd}_{ijkl} \,,
  \label{eq:defTG}
\end{align} 
where $\delta^{abcd}_{ijkl}$ is the generalised Kronecker delta which in four dimensions is equivalent to $\delta^{abcd}_{ijkl} = \varepsilon^{abcd} \varepsilon_{ijkl}$. This term depends on the contortion tensor and its first partial derivatives, and it is quartic in contortion. This is expected as curvature in general is quadratic in contortion, and the Gauss-Bonnet term is itself quadratic in curvature. 

On the other hand, the Gauss-Bonnet boundary term $B_{G}$ reads
\begin{align}
  B_{G} = \frac{1}{e}\delta^{abcd}_{ijkl} \partial_{a}
  \Big[\frac{1}{2}K_{b}{}^{ij}R^{kl}{}_{cd}+K_{b}{}^{ij}K_{c}{}^{k}{}_{f}K_{d}{}^{fl}\Big] \,.
\end{align}
This expression can be deduced from~\cite{Kofinas:2014owa} who do not state this explicitly in four dimensions but provide a general formula using the calculus of forms which can be converted straightforwardly into this form. Equivalently, by using (\ref{relation}), this term can be rewritten depending only on the contorsion tensor as follows
\begin{align}
 B_{G} = \frac{1}{e}\delta^{abcd}_{ijkl} \partial_{a}
 \Big[K_{b}{}^{ij}\Big(K_{c}{}^{kl}{}_{,d}+K_{d}{}^{m}{}_{c}K_{m}{}^{kl}\Big)\Big]\,.
 \end{align}
When discussing the teleparallel equivalent of general relativity we briefly touched upon the issue of Lorentz invariance. As before, it is clear that for instance $T_{G}$ cannot be a Lorentz scalar. To see this, one notes that the final term in the definition~(\ref{eq:defTG}) contains a partial derivative. Therefore this term will contribute second partial derivatives of the local Lorentz transformations which cannot be cancelled by any other term in $T_{G}$. Since the Gauss-Bonnet term $G$ is a Lorentz scalar, these second derivative terms must be cancelled by terms coming from $B_{G}$. Consequently the combination $-T_{G} + B_{G}$ is the unique Lorentz invariant combination which can be constructed. This fact becomes important when considering modified theories of gravity based on the teleparallel equivalent of the Gauss-Bonnet scalar.

In the following we will show some simple examples of the Gauss-Bonnet term and its teleparallel equivalent.

\subsection{Example: FLRW spacetime with diagonal tetrad}

Let us begin with the the FLRW metric and the diagonal tetrad given by
\begin{align}
  e^a_{\mu} = \text{diag}
  \Bigl(
  1,\frac{a(t)}{1+ (k/4)(x^2+y^2+z^2)},\frac{a(t)}{1+ (k/4)(x^2+y^2+z^2)},\frac{a(t)}{1+ (k/4)(x^2+y^2+z^2)}
  \Bigr) \,,
  \label{eq:badflrw}
\end{align}
using spatial Cartensian coordinates. It is straightforward to verify that
\begin{align}
  G &= 24\frac{k}{a^2}\frac{\ddot{a}}{a} + 24\frac{\ddot{a}}{a}\frac{\dot{a}^2}{a^2} \,, \\
  T_{G} &= -24\frac{\ddot{a}}{a}\frac{\dot{a}^2}{a^2} + 2k\frac{k}{a^2}\frac{\ddot{a}}{a}(x^2+y^2+z^2) \,, \\
  B_{G} &= 24\frac{k}{a^2}\frac{\ddot{a}}{a} + 2k\frac{k}{a^2}\frac{\ddot{a}}{a}(x^2+y^2+z^2) \,.
\end{align}
These three quantities display some of the key properties important in this context. Firstly, we note that $T_{G}$ and $B_{G}$ depend on the Euclidean distance from the origin while the Gauss-Bonnet term $G$ is independent of the Cartesian coordinates. The unique linear combination $-T_{G}+B_{G}$ is independent of position. Secondly, in case of a spatially flat universe these terms are absent and the term $B_{G}$ identically vanishes. The terms depending on the spatial coordinates can be changed be working with a different tetrad, or in other words, these terms are affected by local Lorentz transformations. Finding a tetrad for which $T_{G}$ and $B_{G}$ are both independent of the spatial coordinates is a rather involved tasks, however, following the approach outline in \cite{Ferraro:2011us,Tamanini:2012hg} we will show that a tetrad with this property can be constructed. Before doing so, we discuss another example with different symmetry properties. 

\subsection{Example: static spherically symmetric spacetime -- isotropic coordinates}

In this example we consider static and spherically symmetric spacetimes and work with isotropic coordinates $(t,x,y,z)$ to avoid coordinate issues with the tetrads. Including time dependence is straightforward, however, the resulting equations are too involved. We choose
\begin{align}
  e^a_{\mu} = \text{diag}
  \Bigl(
  A(r),B(r),B(r),B(r)
  \Bigr) \,,
  \label{eq:badsss}
\end{align}
where $r=\sqrt{x^2+y^2+z^2}$ is the Euclidean distance from the origin. The metric takes the isotropic form 
\begin{align}
  ds^2 = -A(r)^2 dt^2 + B(r)^2 \bigl(dx^2+dy^2+dz^2\bigr) \,.
\end{align}
The first three quantities of interest $R$, $T$ and $B$ are given by
\begin{align}
  R &= \frac{1}{B^2}
  \Bigl(
  \frac{4}{r}\frac{A'}{A} + \frac{8}{r}\frac{B'}{B} + 2 \frac{A'}{A}\frac{B'}{B} -
  2\frac{B'^2}{B^2} + 2 \frac{A''}{A} + 4 \frac{B''}{B}
  \Bigr) \,, \\
  T &= \frac{1}{B^2}
  \Bigl(
  4 \frac{A'}{A}\frac{B'}{B} + 2\frac{B'^2}{B^2}
  \Bigr) \,, \\
  B &= \frac{1}{B^2}
  \Bigl(
  \frac{4}{r}\frac{A'}{A} + \frac{8}{r}\frac{B'}{B} + 6 \frac{A'}{A}\frac{B'}{B} + 
  2 \frac{A''}{A} + 4 \frac{B''}{B}
  \Bigr) \,.
\end{align}
A direct calculation verifies that indeed $R = -T + B$ as we expected.

Next, we will state the explicit forms of $G$, $T_{G}$ and $B_{G}$ which are
\begin{align}
  G &= \frac{8}{B^4}
  \Bigl( 
  \frac{2}{r^2}\frac{A'}{A}\frac{B'}{B} - \frac{2}{r}\frac{A'}{A}\frac{B'^2}{B^2} -
  3\frac{A'}{A}\frac{B'^3}{B^3} + \frac{2}{r}\frac{A''}{A}\frac{B'}{B} +
  \frac{A''}{A}\frac{B'^2}{B^2} + \frac{2}{r}\frac{A'}{A}\frac{B''}{B} + 
  2\frac{A'}{A}\frac{B' B''}{B^2}
  \Bigr) \,, \\
  T_{G} &=  -\frac{8}{B^4}
  \Bigl( 
  \frac{2}{r}\frac{A'}{A}\frac{B'^2}{B^2} -
  3\frac{A'}{A}\frac{B'^3}{B^3} + \frac{A''}{A}\frac{B'^2}{B^2} + 
  2\frac{A'}{A}\frac{B' B''}{B^2}
  \Bigr) \,, \\
  B_{G} &= \frac{8}{B^4}
  \Bigl( 
  \frac{2}{r^2}\frac{A'}{A}\frac{B'}{B} - \frac{4}{r}\frac{A'}{A}\frac{B'^2}{B^2} + 
  \frac{2}{r}\frac{A''}{A}\frac{B'}{B} + \frac{2}{r}\frac{A'}{A}\frac{B''}{B}
  \Bigr) \,,
\end{align}
which indeed satisfies the required identity
\begin{align}
  G = -T_{G} + B_{G} \,.
\end{align}
We note that the expressions are considerably more complicated than in the previous case. 

The tetrads used in (\ref{eq:badflrw}) and (\ref{eq:badsss}) serve as simple examples which are useful to compute the relevant quantities. However, in the context of extended or modified teleparallel theories of gravity, such tetrads should be avoided. The construction of a suitable static and spherically symmetric tetrad in $f(T)$ gravity, for instance, is rather involved, see~\cite{Ferraro:2011ks}. In general the choice of a suitable parallelisation is a subtle and non-trivial issue, the interested reader is referred to \cite{Ferraro:2014owa}.

\subsection{Example: FLRW spacetime -- good tetrad}

Let us next consider FLRW metric in spherical coordinates given by
\begin{align}
  ds^2 = -dt^2+a(t)^2\Big[\frac{1}{1-kr^2}dr^2+d\Omega^2\Big]\,,
\end{align}
where  $a(t)$ is the scale factor of the universe and $k=\{0,\pm 1\}$ is the spatial curvature which corresponds to flat, close and open cosmologies, respectively.

The simplest tetrad field which yields the above metric is the diagonal 
\begin{align}
  e^{a}_{\mu} = \text{diag}
  \Big(
  1,a(t)/\sqrt{1-kr^2},a(t)r,a(t)r\sin\theta
  \Big)\,.
  \label{eq:tetrad}
\end{align}
However, when this tetrad is used in $f(T)$ gravity it implies an off-diagonal field equation which is highly restrictive, namely the condition $f_{TT}=0$. Such a theory is equivalent to general relativity and hence not a modification. In order to avoid this issue, one can follow the procedure outlined \cite{Ferraro:2011us,Tamanini:2012hg} which allows for the construction of tetrads which result in more favourable field equations. Consider the tetrad (\ref{eq:tetrad}) and perform a general 3-dimensional rotation $\mathcal{R}$ in the tangent space parametrised by three Euler angles $\alpha$, $\beta$, $\gamma$ so that 
\begin{align}
  {\Lambda^a}_b=
  \begin{pmatrix}
    1 & 0 \\
    0 & \mathcal{R}(\alpha,\beta,\gamma) \\
  \end{pmatrix}\,.
  \label{106}
\end{align}
We reduce this transformation considering the following values for the three Euler angles
\begin{align}
  \alpha=\theta-\frac{\pi}{2}\,, \qquad \beta=\phi\,, \qquad \gamma=\gamma(r) \,,
  \label{125}
\end{align}
where $\gamma$ is taken to be a general function of both $t$ and $r$. Doing this means we will work with the rotated tetrad
\begin{align}
  \bar{e}^{a}_{\mu} &= \Lambda^a{}_b e_{\mu}^{b}\,.
  \label{rotatedtetrad}
\end{align}
Next, we focus on the non-flat case $k\neq0$, since the Gauss-Bonnet boundary term $B_{G}=0$ and hence directly $G=-T_{G}$ when $k=0$. By using the rotated tetrad for $k\neq0$, the torsion scalar $T$ and the boundary term $B$  becomes
\begin{align}
  T &= -\frac{4}{a^{2}} \Big(
  \frac{\sqrt{1-k r^2}}{r^2}\Big[r \gamma' \cos \gamma+ \sin \gamma\Big] + \frac{1}{r^2} \Bigr) + 
  6\frac{\dot{a}^2}{a^2} + 2\frac{k}{a^2}
  \,,
  \label{127}\\
  B &= -\frac{4}{a^{2}} \Big(
  \frac{\sqrt{1-k r^2}}{r^2}\Big[r \gamma' \cos \gamma+ \sin \gamma\Big] + \frac{1}{r^2} \Big) +
  6 \frac{\ddot{a}}{a} + 12 \frac{\dot{a}^2}{a^2} + 8 \frac{k}{a^2}
  \,.
  \label{128}
\end{align}  
Here, primes and dots denote derivation with respect to $r$ and $t$, respectively. In order to have $T$ and $B$ position independent we must choose our function $\gamma$ to satisfy
\begin{align}
  \sqrt{1-k r^2} \Big[r \gamma' \cos \gamma+ \sin \gamma\Big] + 1 = 0 \,.
  \label{constraint}
\end{align}

Let us first study the open universe $k=-1$, which from the above equation, give us the following function
\begin{align}
  \gamma(r) = -\arcsin\left[\arcsinh(r)/r \right] \,,
  \label{gamma2}
\end{align}
where we set the constant of integration to zero. Using this choice of $\gamma$ ensures that the first terms in (\ref{127}) and (\ref{128}) disappear thereby making $T$ and $B$ time dependent only. Therefore, the rotated tetrad (\ref{rotatedtetrad}) with $k=-1$ and the function $\gamma$ given by (\ref{gamma2}) is a `good' tetrad in the sense of \cite{Tamanini:2012hg}. Independently of the choice of tetrad we always obtain the usual Ricci scalar
\begin{align}
  R = -T + B = 6 \frac{\ddot{a}}{a} + 6 \frac{\dot{a}^2}{a^2} +6 \frac{k}{a^2} \,.
\end{align}
Moreover, by using this rotated tetrad we find that the Gauss-Bonnet terms are also independent of $r$, one can verify that
\begin{align}
  T_{G} &= -8\frac{\ddot{a}}{a}\Big(3H^2-\frac{1}{a^2}\Big)\,,\\
  B_{G} &= -16\frac{\ddot{a}}{a^3}\,,
\end{align}
and hence the Gauss-Bonnet term in an open universe becomes
\begin{align}
  G = -T_{G} + B_{G} = 24\frac{\ddot{a}}{a}\Big(H^2-\frac{1}{a^2}\Big)\,.
\end{align}

On the other hand, for the closed FLRW universe ($k=+1$), we find that the function $\gamma$ has to be of the form
\begin{align}
  \gamma(r)=-\arcsinh\Big(\sqrt{1+r^2}\Big)\,.
  \label{gammaplus}
\end{align}
This yields
\begin{align}
  T_{G} &= -24\frac{\ddot{a}}{a}\Big(H^2-\frac{1}{a^2}\Big)\,,\\
  B_{G} &=48\frac{\ddot{a}}{a^3}\,,
\end{align}
with the Gauss-Bonnet term given by 
\begin{align}
  G = -T_{G} + B_{G} = 24\frac{\ddot{a}}{a}\Big(H^2+\frac{1}{a^2}\Big)\,.
\end{align}

\section{Modified theories of gravity and their teleparallel equivalents}
\label{sec:ftbgravity}

We are now ready to discuss the general framework of modified theories of gravity and their teleparallel counterparts. In principle our approach could be applied to any metric theory of gravity whose action is based on objects derived from the Riemann curvature tensor. Any such theory can in principle be re-written using the torsion tensor thereby allowing for a teleparallel representation of that same theory.

\subsection{Equations of motion}

We will now consider the framework which includes the teleparallel Gauss-Bonnet and the classical Gauss-Bonnet modified theories of gravity. Inspired by the above discussion, we define the action
\begin{align}
  S = \int 
  \left[ 
    \frac{1}{2\kappa}f(T,B,T_{G},B_{G}) + L_{\rm m}
  \right] e\, d^4x \,,\label{action}
\end{align}
where $f$ is a smooth function of the scalar torsion $T$, the boundary term $B$, the Gauss-Bonnet scalar torsion $T_{G}$ and the boundary Gauss-Bonnet term $B_{G}$.

Variations of the action (\ref{action}) with respect to the tetrad gives
\begin{align}
  \delta S = \int 
  \Big[ 
    \frac{1}{2\kappa}
    \Big(
    f \delta e + 
    e f_{B} \delta B + 
    e f_{T} \delta T + 
    e f_{T_{G}} \delta T_{G} +
    e f_{B_{G}} \delta B_{G}
    \Big) 
    + \delta(e L_{\rm m})
  \Big] 
  \, d^4x \,,
  \label{action2}
\end{align}
where
\begin{align}
  e f_{T} \delta T = {}& -4\Big[
    e(\partial_{\mu}f_{T})S_{a}\,^{\mu\beta}+\partial_{\mu}(e S_{a}\,^{\mu\beta})f_{T}-ef_{T}T^{\sigma}\,_{\mu a}S_{\sigma}\,^{\beta\mu}
    \Big]\delta e^{a}_{\beta} \,,
  \label{deltaT}\\[1ex]
  e f_{B} \delta B = {}& \Big[
    2eE_{a}^{\nu}\nabla^{\beta}\nabla_{\mu}f_{B}-2eE_{a}^{\beta}\Box f_{B}-Bef_{B}E_{a}^{\beta}-4e(\partial_{\mu}f_{B})S_{a}\,^{\mu\beta}
    \Big] \delta e_{\beta}^{a} \,,
  \label{deltaB}\\[1ex]
  e f_{T_{G}} \delta T_{G} = {}& \Big[
    \partial_{\mu}\Big(E_{h}^{\mu}E_{b}^{\beta}(Y^{b}{}_{a}{}^{h}-Y^{h}{}_{a}{}^{b}+Y_{a}{}^{[bh]})\Big)+T^{i}{}_{ab}E^{\beta}_{h}(Y^{b}{}_{i}{}^{h}-Y^{h}{}_{i}{}^{b}+Y_{i}{}^{[bh]}) 
    \nonumber\\
    &-2ef_{T_{G}}\delta^{mbcd}_{ijkl}E_{d}^{\beta}K_{m}{}^{ij}K_{b}{}^{k}{}_{e} \partial_a (K_{c}{}^{el})
    \Big] \delta e_{\beta}^{a}
  \label{deltaTG}\,,\\[1ex]
  ef_{B_{G}}\delta B_{G} = {}& -\Big[\partial_{\mu}\Big((P^{b}{}_{a}{}^{h}-P^{h}{}_{a}{}^{b}+P_{a}{}^{[bh]})E_{h}^{\mu}E_{b}^{\beta}\Big)+T^{i}{}_{ab}E^{\beta}_{h}(P^{b}{}_{i}{}^{h}-P^{h}{}_{i}{}^{b}+P_{i}{}^{[bh]})\nonumber\\
  &-\delta_{ijkl}^{mbcd}eE_{d}^{\beta}\partial_{m}(f_{B_{G}})K_{b}{}^{ij}(\partial_{a}K_{c}{}^{kl})+e\partial_{\mu}(f_{B_{G}})(E_{a}^{\beta}B_{G}^{\mu}-E_{a}^{\mu}B_{G}^{\beta})+ef_{B_{G}}B_{G}E_{a}^{\beta}\Big]\delta e_{\beta}^{a}\,,\label{deltaBG}\\[1ex]
  f \delta e = {}& e f E_{a}^{\beta} \delta e^{a}_{\beta} 
  \label{deltae1} \,.
\end{align}
Here, we introduced the following tensors
\begin{align}
  X^{a}{}_{ij} &= \frac{\partial T_{G}}{\partial K_{a}{}^{ij} }\,, \qquad 
  Y^{b}{}_{ij} = ef_{T_{G}}X^{b}{}_{ij}-2\delta^{cabd}_{elkj} \partial_{\mu} \big(ef_{TG}E^{\mu}_{d}K_{c}{}^{el}K_{a}{}^{k}{}_{i}\big)\,,
  \label{eq:defXY}
\end{align}
and also
\begin{multline}
  P^{b}{}_{ij} = eE_{m}^{\mu}(\partial_{\mu}f_{B_{G}})\Big\{\Big(K_{c}{}^{kl}{}_{,d}+K_{d}{}^{p}{}_{c}K_{p}{}^{kl}\Big)\delta_{ijkl}^{mbcd}+\eta_{pj}\delta_{qckl}^{mdpb}K_{d}{}^{qc}K_{i}{}^{kl}+\delta^{mpcd}_{klij}K_{p}{}^{kl}K_{d}{}^{b}{}_{c}\Big\} \\
  -\delta_{klij}^{acbd}\partial_{\sigma}\Big(eE^{\sigma}_{d}E_{a}^{\mu}(\partial_{\mu}f_{B_{G}})K_{c}{}^{kl}\Big)\,.
\end{multline}
Eqs.~(\ref{deltaT}) and (\ref{deltaB}) were previously derived in \cite{Bahamonde:2015zma} and therefore additional details are suppressed here. On the other hand, the computations of (\ref{deltaTG}) and (\ref{deltaBG}) are rather involved and detailed derivations are given in the Appendices \ref{deltaTGGG} and \ref{deltaBGGG}, respectively. In addition, the explicit form of $X^{a}{}_{ij}$ is showed in Eq.~(\ref{X}) as our definition (\ref{eq:defXY}) hides many of the complications. 

The field equations are very complicated, however, when considering a homogeneous and isotropic spacetime, they simplify substantially and can be presented in closed form. Comparison of these equations with previous results serves as a good consistency check of our calculations.

\subsection{FLRW equations with $k=0$}

For a flat FLRW, the field equations for the $f(T,B,T_{G},B_{G})$ theory are given by
\begin{align}
  f + \frac{6 \dot{a} \dot{f}_{B}}{a} - \frac{12 f_{T} \dot{a}^2}{a^2} -
  \frac{24 \dot{a}^3 \dot{f}_{T_{G}}}{a^3} - \frac{6 f_{B} \left(a \ddot{a}+2 \dot{a}^2\right)}{a^2} +
  \frac{24 f_{T_{G}} \dot{a}^2 \ddot{a}}{a^3} 
  &= 2\kappa\,\rho\,,\\
  f - \frac{4 \dot{a} \dot{f}_{T}}{a} - \frac{8 \dot{a}^2 \ddot{f}_{T_{G}}}{a^2} - 
  \frac{6 f_{B} \left(a\ddot{a}+2\dot{a}^2\right)}{a^2} -
  \frac{4 f_{T} \left(a\ddot{a}+2 \dot{a}^2\right)}{a^2}-
  \frac{16 \dot{a}\ddot{a}\dot{f}_{T_{G}}}{a^2}
  +\frac{24 f_{T_{G}} \dot{a}^2 \ddot{a}}{a^3}+2 \ddot{f}_{B}
  &= -2\kappa\, p\,.
\end{align}	
Here we assumed for simplicity that the matter is a standard perfect fluid with energy density $\rho$ and isotropic pressure $p$. One should make explicit that $\dot{f}_{B}=f_{BB}\dot{B}+f_{BT}\dot{T}+f_{BT_{G}}\dot{T}_{G}+f_{BB_{G}}\dot{B}_{G}$ using the chain rule, so that dot denotes differentiation with respect to cosmic time. It is clear that by setting $f(T,B,T_{G},B_{G})=\mathfrak{f}(R,G)$ the equations for the flat FRWL in $\mathfrak{f}(R,G)$ theory are recovered and are explicitly given by
\begin{align}
  \mathfrak{f}(R,G) + \frac{6 \dot{a}\dot{\mathfrak{f}}_{R} }{a} + \frac{12\mathfrak{f}_{R}\dot{a}^2}{a^2} +
  \frac{24 \dot{a}^3 \dot{\mathfrak{f}}_{G}}{a^3} - 
  \frac{6 \mathfrak{f}_{R} \left(a\ddot{a}+2\dot{a}^2\right)}{a^2} - 
  \frac{24 \mathfrak{f}_{G}\dot{a}^2 \ddot{a}}{a^3} 
  &=2\kappa\, \rho\,,\\
  \mathfrak{f}(R,G) + \frac{4 \dot{a} \dot{\mathfrak{f}}_{R}}{a} + \frac{8 \dot{a}^2 \ddot{\mathfrak{f}}_{G}}{a^2} -
  \frac{2\mathfrak{f}_{R} \left(a \ddot{a}+2 \dot{a}^2\right)}{a^2} + 
  \frac{16 \dot{a}\ddot{a}\dot{\mathfrak{f}}_{G} }{a^2} -
  \frac{24 \mathfrak{f}_{G}\dot{a}^2\ddot{a}}{a^3} + 2\ddot{\mathfrak{f}}_{R} 
  &= -2\kappa\, p\,.
\end{align}

These equations match those reported in \cite{Cognola:2006eg,Kofinas:2014aka} which serves as a good consistency check of our field equations in the teleparallel formulation. 

\subsection{FRWL equations with $k=+1$}

If we consider the rotated tetrad in a closed universe given by Eq.~(\ref{rotatedtetrad}) with $\gamma$ given by (\ref{gammaplus}), the field equations become
\begin{align}
  f + \frac{6 \dot{a} \dot{f}_{B}}{a} - \frac{6 f_{B} \left(a\ddot{a} +2 \dot{a}^2\right)}{a^2} - 
  \frac{12 \dot{a}^2 f_{T}}{a^2} - \frac{48 f_{B_{G}}\ddot{a}}{a^3} + 
  \frac{48 \dot{a} \dot{f}_{B_{G}}}{a^3} -
  \frac{24 \left(\dot{a}^2-1\right) \dot{a} \dot{f}_{T_{G}}}{a^3} +
  \frac{24 f_{T_{G}} \left(\dot{a}^2-1\right) \ddot{a}}{a^3}
  &= 2\kappa\, \rho\,,\\
  f - \frac{48 f_{B_{G}}\ddot{a}}{a^3} - \frac{4 \dot{a} \dot{f}_{T}}{a} +
  \frac{8(1- \dot{a}^2) \ddot{f}_{T_{G}}}{a^2} - \frac{6 f_{B} \left(a \ddot{a}+2 \dot{a}^2\right)}{a^2} -
  \frac{f_{T} \left(4 a \ddot{a}+8 \dot{a}^2-4\right)}{a^2} -
  \frac{16 \dot{a}\ddot{a} \dot{f}_{T_{G}}}{a^2}
  \nonumber \\ +
  \frac{24 f_{T_{G}} \left(\dot{a}^2-1\right) \ddot{a}}{a^3} +
  \frac{16 \ddot{f}_{B_{G}}}{a^2}+2 \ddot{f}_{B}
  &= -2\kappa\, p\,.
\end{align}
By setting $f(T,B,T_{G},B_{G})=\mathfrak{f}(-T+B,-T_{G}+B_{G})=\mathfrak{f}(R,G)$ we formally have $\mathfrak{f}_{G}=\mathfrak{f}_{B_{G}}=-\mathfrak{f}_{T_{G}}$ and $\mathfrak{f}_{R}=\mathfrak{f}_{B}=-\mathfrak{f}_{T}$ and then we recover the usual Gauss-Bonnet equations $\mathfrak{f}(R,G)$ with $k=1$ which are given by 
\begin{align}
  \mathfrak{f}(R,G) + \frac{6 \dot{a} \dot{\mathfrak{f}}_{R} }{a} + \frac{12\mathfrak{f}_{R} \dot{a}^2}{a^2} + 
  \frac{24\dot{a} \left(\dot{a}^2+1\right) \dot{\mathfrak{f}}_{G}}{a^3} - 
  \frac{6 \mathfrak{f}_{R} \left(a\ddot{a}+2\dot{a}^2\right)}{a^2} - 
  \frac{24 \mathfrak{f}_{G} \left(\dot{a}^2+1\right) \ddot{a}}{a^3}
  &= 2\kappa\, \rho\,,\\
  \mathfrak{f}(R,G) + \frac{4 \dot{a} \dot{\mathfrak{f}}_{R}}{a} + \frac{8 \left(\dot{a}^2+1\right) \ddot{\mathfrak{f}}_{G}}{a^2} -
  \frac{2\mathfrak{f}_{R} \left(a\ddot{a}+2 \dot{a}^2+2\right)}{a^2} + \frac{16 \dot{a}\ddot{a}\dot{\mathfrak{f}}_{G} }{a^2} -
  \frac{24 \mathfrak{f}_{G} \left(\dot{a}^2+1\right) \ddot{a}}{a^3}+2 \ddot{\mathfrak{f}}_{R} 
  &= -2\kappa\, p\,.
\end{align}
Finally let us consider the Einstein static universe where all dynamical variables are assumed to be constants which gives
\begin{align}
  \mathfrak{f} = 2\kappa\, \rho_0 \,, \qquad 
  \mathfrak{f} - \mathfrak{f}_R \frac{4}{a_0^2} = -2\kappa\, p_0\,.
\end{align}
For the choice $\mathfrak{f}(R,G) = R + \kappa\, \mathfrak{g}(G)$, we note that $R=6/a_0^2$ and $G=0$ in this case. The field equations reduce simply to
\begin{alignat}{3}
  \frac{6}{a_0^2} + \kappa\, \mathfrak{g}(0) &= 2\kappa\, \rho_0 \,,
  &\qquad &\Leftrightarrow &\qquad
  \frac{3}{a_0^2} &= \kappa\, \rho_0 - \frac{\kappa}{2} \mathfrak{g}(0)\,, \\
  \frac{6}{a_0^2} + \kappa\, \mathfrak{g}(0) - \frac{4}{a_0^2} &= -2\kappa\, p_0\,,
  &\qquad &\Leftrightarrow &\qquad
  -\frac{1}{a_0^2} &= \kappa\, p_0 + \frac{\kappa}{2} \mathfrak{g}(0)\,,
\end{alignat}
which are exactly the $k=+1$ equations reported in \cite{Bohmer:2009fc}, providing us with a second consistency check. It should be noted that the rotated tetrad used for this calculation proves computationally very challenging.

\subsection{Theories with energy-momentum trace}

We will now consider the above framework and include the trace of the energy-momentum tensor to the action (\ref{action}). This gives the extended action
\begin{align}
  S_{\mathcal{T}} = \int 
  \left[ 
    \frac{1}{2\kappa}f(T,B,T_{G},B_{G},\mathcal{T}) + L_{\rm m}
  \right] e\, d^4x \,,
  \label{actiontrace}
\end{align}
where additionally $f$ is a function of the trace of the energy-momentum tensor $\mathcal{T}=E_{a}^{\beta}\mathcal{T}_{\beta}^{a}$. As before $L_{\rm m}$ denotes an arbitrary matter Lagrangian density. We can define the energy-momentum tensor as
\begin{align}
  \mathcal{T}^{a}_{\mu} = \frac{1}{e}\frac{\delta(eL_{m})}{\delta E_{a}^{\mu}}\,.
\end{align}
and assume that the matter Lagrangian only depends explicitly on the tetrads and its derivatives and does not depend on the connection independently. The energy-momentum tensor is then given by
\begin{align}
  \mathcal{T}_{\mu}^{a} = -2e^{a}_{\mu} L_{m} -
  2 \Big(\frac{\partial L_{m}}{\partial E^{\mu}_{a}}\Big) \,.
  \label{emomentumtrace}
\end{align}
Variations of the action (\ref{actiontrace}) with respect to the tetrad gives one additional term, namely
\begin{align}
  e f_{\mathcal{T}} \delta \mathcal{T} &= e f_{\mathcal{T}} (4\Omega_{a}^{\beta}+ \mathcal{T}_{a}^{\beta})\delta e^{a}_{\beta}\label{deltaTTtrace}\,,\\
  \Omega_{a}^{\beta} &= \frac{1}{4}e^{b}_{\alpha}
  \Big(\frac{\delta \mathcal{T}_{b}^{\alpha}}{\delta e_{\beta}^{a}}\Big) =
  \mathcal{T}_{a}^{\beta}+\frac{3}{2}E_{a}^{\beta}L_{m}-\frac{1}{2}e_{\alpha}^{b}
  \Big(\frac{\partial^2 L_{m}}{\partial e_{\beta}^{a}\partial e_{\alpha}^{b}}\Big)\,.
\end{align}
This completes the statement of the field equations. 

\section{Conclusions}

Let us begin these conclusions with a discussion of the relationship between the various modified theories of gravity which are governed by the function $f(T,B,T_{G},B_{G})$. In general these are fourth order theories which violate local Lorentz invariance. Therefore, these theories are quite different from general relativity in many ways. However, for a particular choice of this function, one is able recover general relativity or its teleparallel equivalent. Therein lies the power of this approach, namely one can recover the two equivalent formulations of general relativity using a single unified approach. This in particular clarifies the roles of the total derivative terms present in our framework. As was shown in \cite{Bahamonde:2015zma}, by considering the function $f(T,B)$ one can formulate the teleparallel equivalent of $f(R)$ gravity and identify those parts of the field equations which are part of $f(T)$ gravity, the second order part of the equations which is not locally Lorentz invariant. In analogy to this one can also make the relationships between various modified theories of gravity clear which are based on the Gauss-Bonnet term.

\begin{figure}[!ht]
  \centering
  \includegraphics[width=0.50\textwidth]{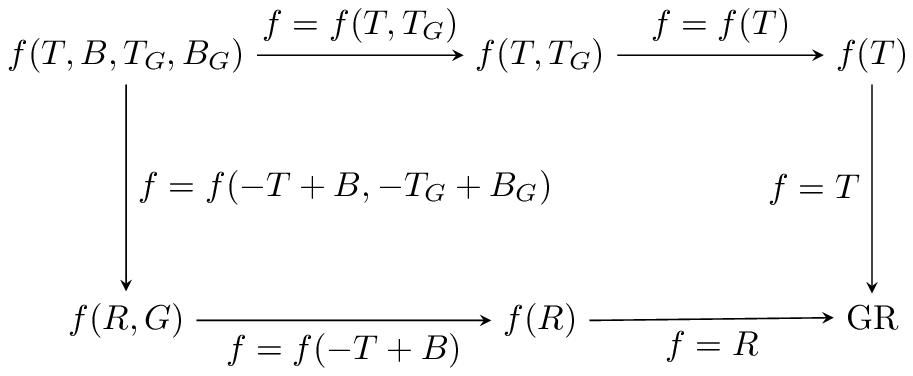}
  \caption{Relationship between different modified gravity models and General Relativity.}
  \label{fig1}
\end{figure} 

The top left corner of Fig.~\ref{fig1} refers to $f(T,B,T_{G},B_{G})$ gravity, the most general theory one can formulate based on the four variables. One can think of the top entries as the teleparallel row and the bottom as the metric row. The arrows indicate the specific choices that have to be made in order to move from one theory to the other. 

\begin{figure}[!b]
  \centering
  \includegraphics[width=0.9\textwidth]{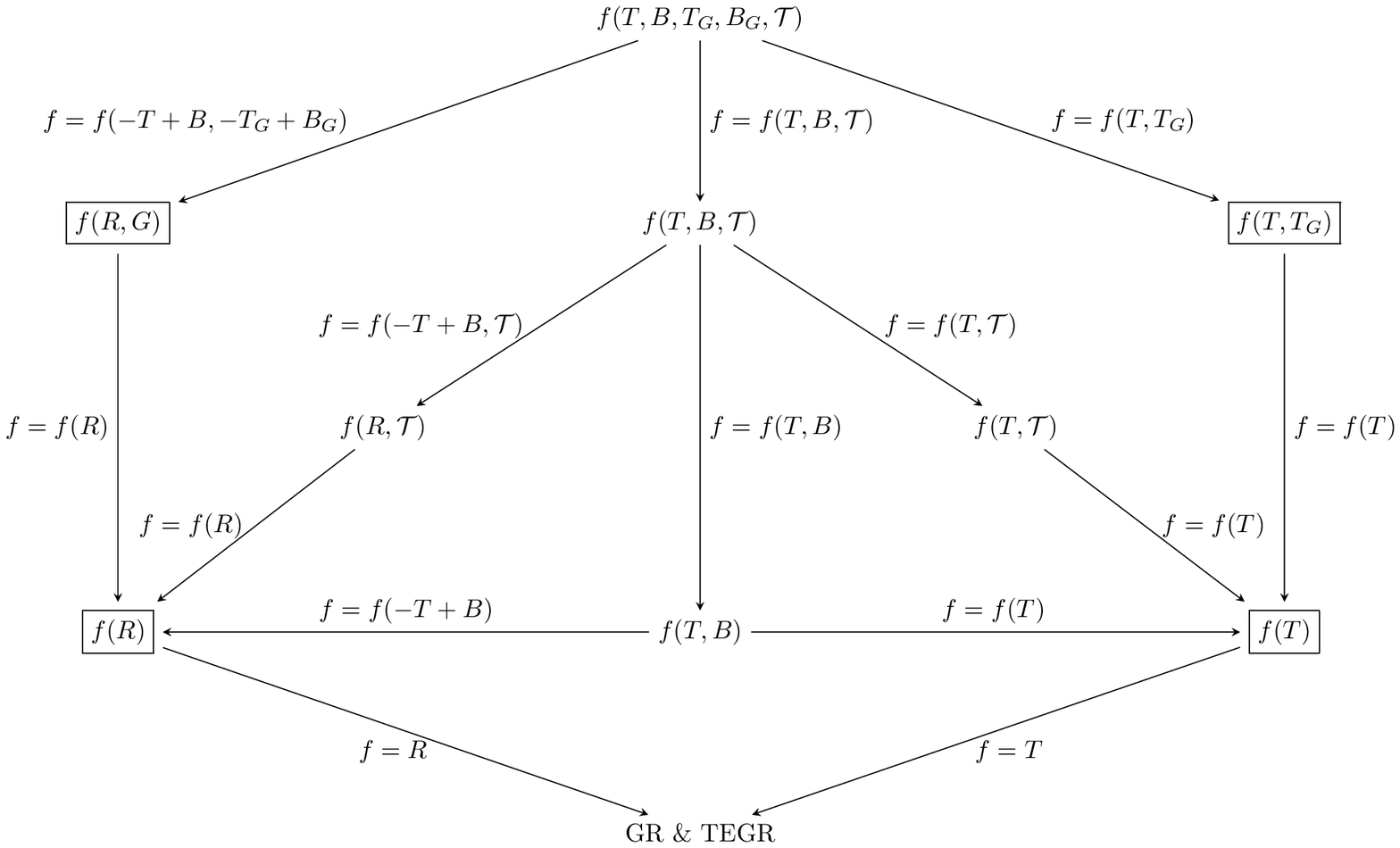}
  \caption{Relationship between different modified gravity models and General Relativity.}
  \label{figtrace}
\end{figure}

If we now include the trace of the energy-momentum tensor to our approach, things get slightly more complicated as the number of possible theories increases quite dramatically. We tried to visualise the entire set of possible theories in Fig.~\ref{figtrace}. 

Now, the left half of the figure corresponds to the metric approach while the right half corresponds to the teleparallel framework. The four main theories of the previous discussions are highlighted by boxes. Many of these theories were considered in isolation in the past and their relationship with other similarly looking theories was only made implicitly. We should also point out that our representation of these theories is only one of the many possibilities and moreover, Fig.~\ref{figtrace} is incomplete. There are many more theories one could potentially construct which we have not mentioned so far. The diagram was constructed having in mind those theories which have been studied in the past.

In constructing the diagram we also made the interesting observation that the theory based on the function $f(R,T)$ should be viewed as a special case of the teleparallel gravity theory $f(T,B)$. To see this, simply recall the principal identity $R=-T+B$ which show that the special choice $f(-T+B,T)$ is the teleparallel equivalent of $f(R,T)$ theory and also that the teleparallel framework should be viewed as the slightly more natural choice for this theory.

We provide a short list of theories in Table \ref{Table1} with accompanying references for the interested reader, mainly focusing on the primary sources or reviews where such theories were considered.

\begin{table}[!ht]
  \centering
  \begin{tabular}{|c|l|}
    \hline
    Theory & Some key references \\
    \hline \hline
    $f(R)$ & Reviews by Sotiriou and Faraoni \cite{Sotiriou:2008rp} \&
    De Felice and Tsujikawa \cite{DeFelice:2010aj} \\
    \hline
    $f(T)$ & Ferraro and Fiorini \cite{Ferraro:2006jd}, and review by Cai et al. \cite{Cai:2015emx} \\
    \hline
    $f(T,B)$ & Bahamonde et al. \cite{Bahamonde:2015zma} \\
    \hline
    $f(R,T)$ & Myrzakulov \cite{Myrzakulov:2012qp} \\
    \hline
    $f(R,G)$ &  Nojiri et al. \cite{Nojiri:2005vv} \\
    \hline
    $f(T,T_{G})$ & Kofinas and Saridakis \cite{Kofinas:2014owa} \& Kofinas et al. \cite{Kofinas:2014aka} \\
    \hline
    $f(R,\mathcal{T})$ & Harko et al. \cite{Harko:2011kv} \\
    \hline
    $f(T,\mathcal{T})$ & Harko et al. \cite{Harko:2014aja} \\
    \hline
  \end{tabular}
  \caption{Short list of previously studied theories covered by the function $f(T,B,T_{G},B_{G},\mathcal{T})$.}
  \label{Table1}
\end{table}

Of the many possible theories one could potentially construct from $f(T,B,T_{G},B_{G},\mathcal{T})$, we identified some which might of interest for future studies. Clearly, there are many theories which do not have a general relativistic counter part like $f(B,T_G,B_B,\mathcal{T})$ since no theory in this class can reduce to general relativity. However, it is always possible to consider such a theory in addition to general relativity by considering for instance a theory based on $-T+f(B,T_{G},B_{G},\mathcal{T})$. For a function linear in its arguments this yields the teleparallel equivalent of general relativity.

It is also useful to make explicit the limitations of the current approach put forward by us. In essence, we are dealing with modified theories of gravity which are based on scalars derived from tensorial quantities of interest, for instance the Ricci scalar or the trace of the energy-momentum tensor. However, theories containing the square of the Ricci tensor or theories containing the term $R_{\mu\nu}\mathcal{T}^{\mu\nu}$ are not currently covered. In principle, it is straightforward though to extend our formalism to such theories. In case of the quantity $R_{\mu\nu}\mathcal{T}^{\mu\nu}$, we would have to recall Eq.~(\ref{riccitensorsplit}) so that this term can be expressed in the teleparallel setting, something that also has not been done yet. Likewise, we could also address quadratic gravity models \cite{Deser:2002jk} which contain squares of the Riemann tensor and use Eq.~(\ref{relation}). Theories depending on higher order derivative terms \cite{Otalora:2016dxe} also require a separate treatment.

The current approach is entirely based on the torsion scalar $T$ which is motivated by its close relation to the Ricci tensor. However, in principle one could follow the work of \cite{Hayashi:1979qx} and decompose the torsion tensor into its three irreducible pieces and construct their respective scalars. This would allow us to study a larger class of models based on those three scalars and the boundary term. To the best of our knowledge this has not been considered in the past and would make an interesting further development. 

For many years now, an ever increasing number of modifications of general relativity has been considered. In this work we focused only on theories where the gravitational field can either be modelled using the metric of the tetrad. Hence, we excluded all types of metric affine theories where the metric and the torsion tensor are treated as two independent dynamical variables. It would be almost impossible to present a visualisation that encompasses all those theories as well. Even this would represent only a fraction of what is referred to as modified gravity. It would still exclude higher dimensional models, Einstein-Aether models, Ho\v{r}ava-Lifshitz theory and many others. It is also interesting to note that $f(R)$ gravity for instance can be formulated as a theory based on a non-minimally coupled scalar field. Hence, many of the theories in Fig.~\ref{figtrace} might also have various other representations which in turn might be connected in different manners.

This discussion motivates the process of classifying the different families of modifications of general relativity and their possible interrelations, followed by a broad investigation of which theories should not be studied further due to incompatibilities with well established observational bounds. It appears that we are reaching the point where we possibly do not need more theories but rather an improved sense of direction for future developments. 

\begin{acknowledgments}
We would like to thank Franco Fiorini and Emmanuel Saridakis for their valuable comments on this manuscript. SB is supported by the Comisi{\'o}n Nacional de Investigaci{\'o}n Cient{\'{\i}}fica y Tecnol{\'o}gica (Becas Chile Grant No.~72150066). This article is partly based upon work from COST Action CA15117 (Cosmology and Astrophysics Network for Theoretical Advances and Training Actions), supported by COST (European Cooperation in Science and Technology).
\end{acknowledgments}

\appendix

\section{Derivation of the field equations}

\subsection{Variation of $B_{G}$}
\label{deltaBGGG}

The Gauss-Bonnet boundary term is given by
\begin{align}
	B_{G} = \frac{1}{e}\epsilon_{ijkl}\epsilon^{bcda}\partial_{a}\Big[\frac{1}{2}K_{b}{}^{ij}R^{kl}{}_{cd}+K_{b}{}^{ij}K_{c}{}^{k}{}_{f}K_{d}{}^{fl}\Big]\,,
\end{align}
or equivalently
\begin{align} 
  B_{G} = \frac{1}{e}\partial_{\mu}(eE_{a}^{\mu}B_{G}^{a})\,,
\end{align}
where we introduced the vector $B_{G}^a$ by
\begin{align}
  B_{G}^{a} = \epsilon_{ijkl}\epsilon^{bcda}\Big(\frac{1}{2}K_{b}{}^{ij}R^{kl}{}_{cd}+K_{b}{}^{ij}K_{c}{}^{k}{}_{f}K_{d}{}^{fl}\Big)\,.
  \label{eq:app_ba}
\end{align}
Using the relationship between the contortion tensor and the Riemann tensor, this means Eq.~(\ref{relation}), and recalling that in four dimensions $\epsilon_{ijkl}\epsilon^{bcd\mu}=\delta_{ijkl}^{bcd\mu}$, the above term (\ref{eq:app_ba}) can be rewritten as
\begin{align} 
  B_{G}^{a} = \delta_{ijkl}^{abcd}K_{b}{}^{ij}\Big((K_{c}{}^{kl})_{,d}+K_{d}{}^{m}{}_{c}K_{m}{}^{kl}\Big)\,.\label{BGa}
\end{align}

We are now considering variations of the function $f(T,B,T_{G},B_{G})$ with respect to the tetrad fields, beginning with the quantity $B_{G}$, which yields
\begin{align}
  ef_{B_{G}} \delta B_{G} = \Big[e\partial_{\mu}(f_{B_{G}})(E_{a}^{\mu}B_{G}^{\beta}-E_{a}^{\beta}B_{G}^{\mu})-ef_{B_{G}}B_{G}E_{a}^{\beta}\Big]\delta e ^{a}_{\beta} - eE_{a}^{\mu}\partial_{\mu}(f_{B_{G}})\delta B_{G}^{a}\,,
	\label{deltaBG2}
\end{align}
where $f_{B_{G}}=\partial f(T,B,T_{G},B_{G})/\partial B_{G}$. We used $\delta E^{\sigma}_{m}=-E^{\sigma}_{n}E^{\mu}_{m}\delta e^{n}_{\mu}$ and $\delta e = eE_{a}^{\beta}\delta e_{\beta}^{a}$, and boundary terms were neglected. The final term in the above equation reads
\begin{align}
  eE_{a}^{\mu}\partial_{\mu}(f_{B_{G}})\delta B_{G}^{a} = P^{b}{}_{ij}\delta K_{b}{}^{ij}-\delta_{ijkl}^{mbcd}eE_{d}^{\beta}\partial_{m}(f_{B_{G}})K_{b}{}^{ij}(\partial_{a}K_{c}{}^{kl})\delta e^{a}_{\beta}\,,
  \label{BGdelta}
\end{align}
where again we neglected boundary terms and for simplicity we introduced the following tensor
\begin{align}
  P^{b}{}_{ij} &= eE_{m}^{\mu}\partial_{\mu}(f_{B_{G}})\Big\{\Big((K_{c}{}^{kl})_{,d}+K_{d}{}^{p}{}_{c}K_{p}{}^{kl}\Big)\delta_{ijkl}^{mbcd}+\eta_{pj}\delta_{qckl}^{mdpb}K_{d}{}^{qc}K_{i}{}^{kl}+\delta^{mpcd}_{klij}K_{p}{}^{kl}K_{d}{}^{b}{}_{c}\Big\}
  \nonumber\\
  & -\delta_{klij}^{acbd}\partial_{\sigma}\Big(eE^{\sigma}_{d}E_{a}^{\mu}\partial_{\mu}(f_{B_{G}})K_{c}{}^{kl}\Big)\,.
  \label{PP}
\end{align}
We take note of $\delta K_{b}{}^{ij}$ in Eq.~(\ref{BGdelta}) which needs to be expressed as a variation with respect to the tetrad $\delta e^{a}_{\beta}$. Therefore, we firstly compute how an arbitrary tensor $D^{b}{}_{ij}\delta K_{b}{}^{ij}$ changes its form in this context. This formula will be useful for computing $P^{b}{}_{ij}\delta K_{b}{}^{ij}$ and is also  needed when computing the variations of  $T_{G}$ in the second part of this appendix.

Recall the contortion and torsion tensors, respectively
\begin{align}
  K_{b}{}^{ij} &= \frac{1}{2}\Big(T^{i}{}_{b}{}^{j}-T^{j}{}_{b}{}^{i}+T_{b}{}^{ij}\Big)\,,
  \label{Klatin}\\
  T^{i}{}_{bh} &= E_{b}^{\mu}E_{h}^{\nu}\Big(\partial_{\mu}e_{\nu}^{i}-\partial_{\nu}e_{\mu}^{i}\Big)\,.
  \label{Tlatin}
\end{align}
Beginning with (\ref{Klatin}) we have that
\begin{align}
  D^{b}{}_{ij}\delta K_{b}{}^{ij} =
  D^{b}{}_{[ij]}\delta K_{b}{}^{[ij]} =
  \frac{1}{2}\Big(D^{b}{}_{i}{}^{h}-D^{h}{}_{i}{}^{b}+D_{i}{}^{[bh]}\Big)\delta T^{i}{}_{bh}=
  \frac{1}{2}C_{i}{}^{bh}\delta T^{i}{}_{bh}\,,
\end{align}
where for simplicity we have introduced the tensor
\begin{align}
  C_{i}{}^{bh} = D^{b}{}_{i}{}^{h} - D^{h}{}_{i}{}^{b} + D_{i}{}^{[bh]} = -C_{i}{}^{hb}\,.
  \label{C}
\end{align}
This tensor needs to be skew-symmetric in its last two indices since $\delta T^{i}{}_{bh}$ is skew-symmetric in this pair.

Next, by using (\ref{Tlatin}) and neglecting boundary terms we find 
\begin{align}
  D^{b}{}_{ij}\delta K_{b}{}^{ij} =
  \Big[
    \partial_{\mu} \Big(C_{a}{}^{bh}E_{h}^{\mu}E_{b}^{\beta}\Big) +T^{i}{}_{ab}E^{\beta}_{h}C_{i}{}^{bh}
    \Big]
  \delta e_{\beta}^{a}
  \label{DdeltaK}\,.
\end{align}
Equivalently, by using (\ref{C}) we find explicitly that for any specific tensor $D_{i}{}^{bh}$ the transformation from $D^{b}{}_{ij}\delta K_{b}{}^{ij}$ to terms with $\delta e_{\beta}^{a}$ will be
\begin{align}
  D^{b}{}_{ij}\delta K_{b}{}^{ij} =
  \Big[
    \partial_{\mu}\Big((D^{b}{}_{a}{}^{h}-D^{h}{}_{a}{}^{b}+D_{a}{}^{[bh]})E_{h}^{\mu}E_{b}^{\beta}\Big)+
    T^{i}{}_{ab}E^{\beta}_{h}(D^{b}{}_{i}{}^{h}-D^{h}{}_{i}{}^{b}+D_{i}{}^{[bh]})
  \Big]
  \delta e_{\beta}^{a}
  \label{DdeltaK1}\,.
\end{align}

Now, if we change $D^{b}{}_{ij} \rightarrow P^{b}{}_{ij}$ we find the useful equation
\begin{align}
  P^{b}{}_{ij}\delta K_{b}{}^{ij} =
  \Big[
    \partial_{\mu}\Big((P^{b}{}_{a}{}^{h}-P^{h}{}_{a}{}^{b}+P_{a}{}^{[bh]})E_{h}^{\mu}E_{b}^{\beta}\Big)+
    T^{i}{}_{ab}E^{\beta}_{h}(P^{b}{}_{i}{}^{h}-P^{h}{}_{i}{}^{b}+P_{i}{}^{[bh]})
  \Big]
  \delta e_{\beta}^{a}\,.
  \label{PdeltaKfinal}
\end{align}

Finally, if we replace (\ref{PdeltaKfinal}) in (\ref{BGdelta}) and then replace that expression in (\ref{deltaBG2}) we find the variations of the Gauss-Bonnet boundary term with respect to the tetrad. This is given by
\begin{align}
  ef_{B_{G}}\delta B_{G} = {}&
  -\Big[
    \partial_{\mu}\Big((P^{b}{}_{a}{}^{h}-P^{h}{}_{a}{}^{b}+P_{a}{}^{[bh]})E_{h}^{\mu}E_{b}^{\beta}\Big)+
    T^{i}{}_{ab}E^{\beta}_{h}(P^{b}{}_{i}{}^{h}-P^{h}{}_{i}{}^{b}+P_{i}{}^{[bh]})
    \nonumber\\
    {}& -\delta_{ijkl}^{mbcd}eE_{d}^{\beta}\partial_{m}(f_{B_{G}})K_{b}{}^{ij} K_{c}{}^{kl}{}_{,a}+e\partial_{\mu}(f_{B_{G}})(E_{a}^{\beta}B_{G}^{\mu}-E_{a}^{\mu}B_{G}^{\beta})+ef_{B_{G}}B_{G}E_{a}^{\beta}\Big]\delta e_{\beta}^{a}\,,
\end{align}
where $P^{b}{}_{ij}$ is explicitly given by Eq.~(\ref{PP}).

\subsection{Variation of $T_{G}$ }
\label{deltaTGGG}

For simplicity, we will split $T_{G}$ in four parts as follows
\begin{align}
  T_{G} &= (K_{a}{}^{i}{}_{e}K_{b}{}^{ej}K_{c}{}^{k}{}_{f}K_{d}{}^{fl}-2K_{a}{}^{ij}K_{b}{}^{k}{}_{e}K_{c}{}^{e}{}_{f}K_{d}{}^{fl} \,+2K_{a}{}^{ij}K_{b}{}^{k}{}_{e}K_{f}{}^{el}K_{d}{}^{f}{}_{c}+2K_{a}{}^{ij}K_{b}{}^{k}{}_{e}K_{c}{}^{el}{}_{,d})\delta^{abcd}_{ijkl}\,,
  \nonumber \\
  &= T_{G1}+T_{G2}+T_{G3}+T_{G4}\,,
\end{align}
where $T_{G1}$, $T_{G2}$, $T_{G3}$ and $T_{G4}$ are the first, second, third and fourth term of the right-hand sides, respectively. Variations of the $T_{G(i)}$, $i=1,2,3,4$ contributions with respect to the tetrad can be expressed as
\begin{align}
  ef_{T_{G}}\delta T_{G} = ef_{T_{G}}(\delta T_{G1}+\delta T_{G2}+\delta T_{G3}+\delta T_{G4})\,.\label{efdeltaTG0}
\end{align}
Here, $f_{T_{G}}$ stands for the partial derivative of $f(T,B,T_{G},T_{B})$ with respect to $T_{G}$. The first, second and third term can be computed without difficulty, yielding
\begin{align}
  \delta T_{G1} &= \Big[K_{bj}{}^{e}K_{c}{}^{k}{}_{f}K_{d}{}^{fl}\delta^{abcd}_{iekl}+K_{b}{}^{e}{}_{i}K_{c}{}^{k}{}_{f}K_{d}{}^{fl}\delta^{bacd}_{ejkl}+K_{c}{}^{k}{}_{e}K_{b}{}^{ef}K_{dj}{}^{l}\delta^{cbad}_{kfil}+K_{d}{}^{f}{}_{e}K_{b}{}^{el}K_{c}{}^{k}{}_{i}\delta^{dbca}_{flkj}\Big]\delta K_{a}{}^{ij}\,,
  \label{TG1}\\
  \delta T_{G2} &= -2\Big[K_{b}{}^{k}{}_{e}K_{c}{}^{e}{}_{f}K_{d}{}^{fl}\delta^{abcd}_{ijkl}+K_{b}{}^{ke}K_{cjf}K_{d}{}^{fl}\delta^{bacd}_{keil}+K_{c}{}^{ef}K_{b}{}^{k}{}_{i}K_{dj}{}^{l}\delta^{cbad}_{efkl}+K_{d}{}^{fl}K_{b}{}^{k}{}_{e}K_{c}{}^{e}{}_{i}\delta^{dbca}_{flkj}\Big]\delta K_{a}{}^{ij}\,,
  \label{TG2}\\
  \delta T_{G3} &= 2\Big[ K_{b}{}^{k}{}_{e}K_{f}{}^{el}K_{d}{}^{f}{}_{c}\delta^{abcd}_{ijkl}+K_{b}{}^{ke}K_{fj}{}^{l}K_{d}{}^{f}{}_{c}\delta^{bacd}_{keil}+K_{f}{}^{el}K_{b}{}^{k}{}_{i}K_{d}{}^{a}{}_{c}\delta^{fbcd}_{elkj}+K_{d}{}^{fm}K_{b}{}^{k}{}_{e}K_{i}{}^{el}\eta_{jc}\delta^{dbca}_{fmkl}\Big]\delta K_{a}{}^{ij}\,.
  \label{TG3}
\end{align}
For the final term $e f_{T_{G}} \delta T_{G4}$ we need to be careful since we need to integrate by parts and hence need to change $\partial_{d}$ to $\partial_{d}=E^{\mu}_{d}\partial_{\mu}$. Therefore, we need to compute the following term
\begin{align}
  ef_{T_{G}}\delta T_{G4} = 2ef_{T_{G}}\delta
  \Big[
    K_{a}{}^{ij}K_{b}{}^{k}{}_{e}K_{c}{}^{el}{}_{,d}
    \Big] \delta^{abcd}_{ijkl} =
  2 ef_{TG} \delta
  \Big[
    E_{d}^{\mu}K_{a}{}^{ij}K_{b}{}^{k}{}_{e} \partial_{\mu} (K_{c}{}^{el})
    \Big] \delta^{abcd}_{ijkl}\,.
\end{align}

By ignoring boundary terms, these terms become
\begin{align}
  ef_{TG}\delta T_{G4} = {}& 2
  \Big[ef_{T_{G}}K_{b}{}^{k}{}_{e}K_{c}{}^{el}{}_{,d}\delta^{abcd}_{ijkl}+ef_{T_{G}}K_{b}{}^{ke}K_{cj}{}^{l}{}_{,d}\delta^{bacd}_{keil}-\delta^{cbad}_{elkj} \partial_{\mu}
    \Big(eE^{\mu}_{d}f_{T_{G}}K_{c}{}^{el}K_{b}{}^{k}{}_{i}\Big)
    \Big]\delta K_{a}{}^{ij}
  \nonumber\\
      {}& -2ef_{T_{G}}\delta^{mbcd}_{ijkl}E_{d}^{\beta}E^{\mu}_{a}K_{m}{}^{ij}K_{b}{}^{k}{}_{e} \partial_{\mu} (K_{c}{}^{el})\delta e_{\beta}^{a} \,.
      \label{TG4}
\end{align}

Now, by adding (\ref{TG1})-(\ref{TG3}) and (\ref{TG4}) we find
\begin{align}
  ef_{T_{G}} \delta T_{G} =
  \Big[
    ef_{T_{G}}X^{a}{}_{ij}-2\delta^{cbad}_{elkj} \partial_{\mu}
    \Big(eE^{\mu}_{d}f_{T_{G}}K_{c}{}^{el}K_{b}{}^{k}{}_{i}\Big)
    \Big] \delta K_{a}{}^{ij} -
  2ef_{T_{G}}\delta^{mbcd}_{ijkl}E_{d}^{\beta}K_{m}{}^{ij}K_{b}{}^{k}{}_{e}K_{c}{}^{el}{}_{,a}\delta e_{\beta}^{a}\,,
  \label{efdeltaTG}
\end{align}
where we have introduced the following tensor
\begin{align}
  X^{a}{}_{ij} = {}& K_{bj}{}^{e}K_{c}{}^{k}{}_{f}K_{d}{}^{fl}\delta^{abcd}_{iekl}+K_{b}{}^{e}{}_{i}K_{c}{}^{k}{}_{f}K_{d}{}^{fl}\delta^{bacd}_{ejkl}+K_{c}{}^{k}{}_{e}K_{b}{}^{ef}K_{dj}{}^{l}\delta^{cbad}_{kfil}+K_{d}{}^{f}{}_{e}K_{b}{}^{el}K_{c}{}^{k}{}_{i}\delta^{dbca}_{flkj}
  \nonumber\\
      {}& -2K_{b}{}^{k}{}_{e}K_{c}{}^{e}{}_{f}K_{d}{}^{fl}\delta^{abcd}_{ijkl}-2K_{b}{}^{ke}K_{cjf}K_{d}{}^{fl}\delta^{bacd}_{keil}-2K_{c}{}^{ef}K_{b}{}^{k}{}_{i}K_{dj}{}^{l}\delta^{cbad}_{efkl}-2K_{d}{}^{fl}K_{b}{}^{k}{}_{e}K_{c}{}^{e}{}_{i}\delta^{dbca}_{flkj}
  \nonumber\\
      {}& +2K_{b}{}^{k}{}_{e}K_{f}{}^{el}K_{d}{}^{f}{}_{c}\delta^{abcd}_{ijkl}+2K_{b}{}^{ke}K_{fj}{}^{l}K_{d}{}^{f}{}_{c}\delta^{bacd}_{keil}+2K_{f}{}^{el}K_{b}{}^{k}{}_{i}K_{d}{}^{a}{}_{c}\delta^{fbcd}_{elkj}+2K_{d}{}^{fc}K_{b}{}^{k}{}_{e}K_{i}{}^{el}\eta_{mj}\delta^{dbma}_{fckl}
  \nonumber\\
      {}& +2K_{b}{}^{k}{}_{e}K_{c}{}^{el}{}_{,d}\delta^{abcd}_{ijkl}+2K_{b}{}^{ke}K_{cj}{}^{l}{}_{,d}\delta^{bacd}_{keil}\,.
      \label{X}
\end{align}
It can be shown easily that this long expression is also equivalent to
\begin{align}
  X^{a}{}_{ij} = \frac{\partial T_{G}}{\partial K_{a}{}^{ij}} =
  \frac{\partial T_{G1}}{\partial K_{a}{}^{ij} } +
  \frac{\partial T_{G2}}{\partial K_{a}{}^{ij} } +
  \frac{\partial T_{G3}}{\partial K_{a}{}^{ij} } +
  2\delta^{fbcd}_{mnkl}K_{c}{}^{el}{}_{,d}\frac{\partial }{\partial K_{a}{}^{ij} }
  \Big[K_{f}{}^{mn}K_{b}{}^{k}{}_{e}\Big]\,.
\end{align}
Next, for simplicity we will introduce the tensor 
\begin{align}
  Y^{b}{}_{ij} = ef_{T_{G}}X^{b}{}_{ij}-2\delta^{cabd}_{elkj}
  \partial_{\mu} \Big(ef_{TG}E^{\mu}_{d}K_{c}{}^{el}K_{a}{}^{k}{}_{i}\Big)\,,
  \label{Y}
\end{align}
to rewrite Eq.~(\ref{efdeltaTG}) as 
\begin{align}
  ef_{T_{G}} \delta T_{G} = Y^{b}{}_{ij} \delta K_{b}{}^{ij}-2ef_{T_{G}}\delta^{mbcd}_{ijkl}E_{d}^{\beta}K_{m}{}^{ij}K_{b}{}^{k}{}_{e} K_{c}{}^{el}{}_{,a} \delta e_{\beta}^{a}\,.
\end{align}
Finally, by using equation (\ref{DdeltaK1}), we can change $\delta K_{a}{}^{ij}$ to $\delta e_{\beta}^{a}$ by changing $D^{a}_{ij}$ to $Y^{a}_{ij}$. Doing that, we finally find that the variations with respect to the $T_{G}$ part is
\begin{eqnarray}
ef_{T_{G}} \delta T_{G}&=&\Big[\partial_{\mu}\Big((Y^{b}{}_{a}{}^{h}-Y^{h}{}_{a}{}^{b}+Y_{a}{}^{[bh]})E_{h}^{\mu}E_{b}^{\beta}\Big)+T^{i}{}_{ab}E^{\beta}_{h}(Y^{b}{}_{i}{}^{h}-Y^{h}{}_{i}{}^{b}+Y_{i}{}^{[bh]})\nonumber\\
&&-2ef_{T_{G}}\delta^{mbcd}_{ijkl}E_{d}^{\beta}K_{m}{}^{ij}K_{b}{}^{k}{}_{e}K_{c}{}^{el}{}_{,a}\Big]\delta e_{\beta}^{a}\,,
\end{eqnarray}
where $Y^{b}{}_{ij}$ is explicitly given by Eq. (\ref{Y}). 

\end{document}